\documentclass[journal]{IEEEtran}

\usepackage{comment}
\renewcommand{\baselinestretch}{0.92}
\usepackage{changepage}
\usepackage{amsmath}
\usepackage{amssymb}
\usepackage[dvips]{graphicx}
\usepackage{setspace}
\usepackage{epsfig}
\usepackage{color}
\usepackage{morefloats}
\usepackage{enumerate}
\usepackage{bbm}
\usepackage{cases}
\usepackage{enumitem}

\usepackage{graphicx}

\allowdisplaybreaks

\usepackage{amsfonts} 
\usepackage{bm} 
\usepackage{comment}
\usepackage{caption}
\usepackage{subcaption}

\usepackage{float}        
\usepackage{stfloats} 

\usepackage[a4paper,
            bindingoffset=0.2in,
            left=0.6in,
            right=0.6in,
            top=0.7in,
            bottom=0.9in,
            footskip=.25in]{geometry}
\begin{document}
\renewcommand{\baselinestretch}{0.85}

\title{Bridging Simulation and Reality: A 3D Clustering-Based Deep Learning Model for UAV-Based RF Source Localization}
\author{Ismail Guvenc\vspace{-0.5cm}}
\author{Saad Masrur and \.{I}smail G\"{u}ven\c{c}\\
Department of Electrical and Computer Engineering, North Carolina State University, Raleigh, NC\\
{\tt \{smasrur,iguvenc\}@ncsu.edu}

\thanks{This work is supported by the NSF under the award numbers CNS-1939334 and CNS-2332834.}
}




\maketitle
\thispagestyle{empty}
\begin{abstract}
\pagestyle{empty}
Localization of radio frequency (RF) sources has critical applications, including search and rescue, jammer detection, and monitoring of hostile activities. Unmanned aerial vehicles (UAVs) offer significant advantages for RF source localization (RFSL) over terrestrial methods, leveraging autonomous 3D navigation and improved signal capture at higher altitudes. Recent advancements in deep learning (DL) have further enhanced localization accuracy, particularly for outdoor scenarios. DL models often face challenges in real-world performance, as they are typically trained on simulated datasets that fail to replicate real-world conditions fully. To address this, we first propose the \emph{Enhanced Two-Ray} propagation model, reducing the simulation-to-reality gap by improving the accuracy of propagation environment modeling. For RFSL, we propose the \emph{3D Cluster-Based RealAdaptRNet}, a DL-based method leveraging 3D clustering-based feature extraction for robust localization. Experimental results demonstrate that the proposed \emph{Enhanced Two-Ray} model provides superior accuracy in simulating real-world propagation scenarios compared to conventional free-space and two-ray models. Notably, the \emph{3D Cluster-Based RealAdaptRNet}, trained entirely on simulated datasets, achieves exceptional performance when validated in real-world environments using the AERPAW physical testbed, with an average localization error of $18.2$~m. The proposed approach is computationally efficient, utilizing $33.5\times$ fewer parameters, and demonstrates strong generalization capabilities across diverse trajectories, making it highly suitable for real-world applications.

\pagestyle{empty}

\textit{Index~Terms}--- 3D Clustering, AERPAW, Deep Learning, Digital Twin, Localization, Real-World Testing, Simulation-to-Reality, UAV.
\end{abstract}




\vspace{-0.3cm}
\section{Introduction}\label{sec:intro}
\pagestyle{empty}

Unmanned aerial vehicles (UAVs) have become essential tools for applications such as search and rescue, surveillance, and radio frequency (RF) source localization \cite{kwon2023rf}. Their flexibility, cost-effectiveness, and autonomous 3D navigation capabilities make them well-suited for enhancing coverage and efficiency in localization tasks \cite{10605607}. Accurately modeling the air-to-ground (A2G) wireless channel is crucial for advancing UAV-assisted sensing and localization.

Various studies have explored RF source localization (RFSL) using UAVs, employing conventional algorithms such as linear least squares and particle filters \cite{kwon2023rf}, \cite{hasanzade2018rf}. 
The impact of transmitter and receiver antenna patterns on localization accuracy is analyzed in \cite{sinha2022impact}, while their effect on A2G propagation models is quantified in \cite{maeng2023impact}. 


Most UAV-based localization studies use the free-space path loss model for outdoor propagation due to its simplicity 
\cite{kwon2023rf}. Traditional free-space and two-ray propagation models \cite{maeng20233d}, while simple, fail to capture the complexities of real-world environments, leading to a gap between simulated and real-world performance. Received signal strength (RSS) at the UAV is significantly influenced by factors such as dynamic 3D propagation conditions, UAV movements (roll, pitch, and yaw), the 3D radiation pattern of the antenna, multipath effects, and shadowing caused by the UAV's structure. These factors can cause power variations of up to 60 dB for narrow-band links, making accurate modeling essential. In \cite{10000615}, the authors investigated the impact of UAV movements on the temporal autocorrelation function (ACF) of the A2G wireless channel, modeling the UAV movements as positional variations of the UAV but neglecting their effect on antenna gains.


To address these challenges, we propose an enhanced two-ray propagation model that incorporates UAV dynamics and structural shadowing to bridge the simulation-to-reality (sim2real) gap. While fingerprint-based localization using deep learning (DL) has seen significant advancements, particularly for outdoor applications \cite{abubakr2023novel}, \cite{asaad2022comprehensive}, most existing DL-based localization solutions are trained and evaluated solely on simulated datasets. Such approaches often fail to perform effectively in real-world scenarios due to the inherent disparities between simulated and real-world environments.
Recognizing this critical limitation, we introduce the \emph{3D Cluster-Based RealAdaptRNet}, a model specifically designed to generalize from simulated training to real-world deployments. The model is trained entirely on simulated datasets for scalability and computational efficiency, but its performance is rigorously validated using real-world data collected on the Aerial Experimentation and Research Platform for Advanced Wireless (AERPAW)  physical testbed. Unlike existing methods, our approach explicitly addresses the sim2real gap, demonstrating that DL models trained on simulations can achieve exceptional accuracy and generalization when tested in real-world environments.



\vspace{-0.1cm}

\begin{figure}
	\includegraphics[width=0.8\linewidth]{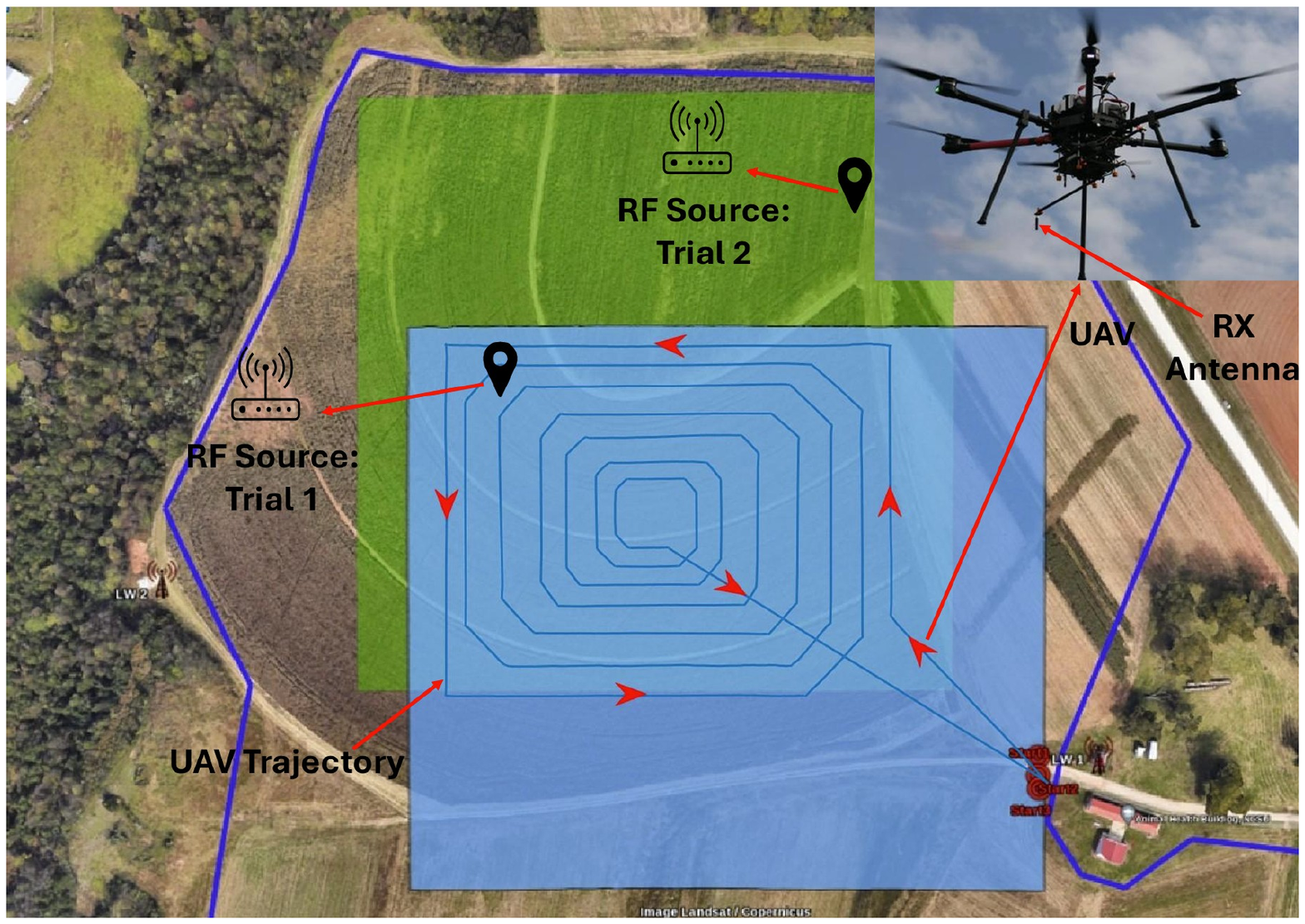}
	\centering
	\caption{Secnario in AERPAW AFAR challenge Lake Wheeler testbed setup for RFSL. The blue region indicates the permissible UAV flight area, while the RF source can be located anywhere within the green-shaded region. The figure also depicts the fixed-way-points spiral trajectory designed to sweep the area and collect RSS measurements for localization. The upper-right corner displays the UAV used in the experiment.}
	\label{Sys}
\vspace{-0.5cm}
\end{figure} 

\section{System Model}\label{sec:system}
The AERPAW \cite{gurses2024digital} is a state-of-the-art testbed for wireless research using UAVs. In Fall 2023, AERPAW hosted the AERPAW Find a Rover (AFAR) competition, which challenged participants to design and implement algorithms for UAVs equipped with wireless receivers to localize a hidden ground-based RF source transmitting a narrowband signal with a 12 kHz bandwidth. The algorithms were developed in the Digital Twin of AERPAW \cite{gurses2024digital}. Several teams contributed innovative localization approaches and algorithms during the competition~\cite{kudyba2024uav}, \cite{aerpaw_afarchallenge}.
Building on the AFAR competition's objectives, this paper proposes a robust solution for accurate ground-based RFSL. The RF source position, \( P_{\rm RF} = (x, y, z) \), lies within the green region shown in Fig. \ref{Sys}, while the UAV operates within the blue-shaded region in Fig. \ref{Sys} for a duration of $T = 10$ minutes.

The UAV, serving as the receiver, measures RSS at an interval of 0.03 seconds. Consequently, the total number of received samples can be approximated as \( S = \frac{T \times 60}{0.03} \). The discretized position of the UAV for the \( s \)-th collected sample is represented as \( \mathbf{\ell}_{\text{UAV}, s} = (r_{ \text{x}, s}, r_{\text{y}, s}, r_{\text{z}, s}) \), where \( s \in \{1, 2, \ldots, S\} \). The corresponding RSS at this position is denoted by \( p_{\text{r}, s} \). Over the complete UAV flight duration \( T \), the set of RSS values collected is represented as \( \mathbf{p}_{\text{r}} \in \mathbb{R}^{S \times 1} \). The associated \( x \)- and \( y \)-coordinates of the UAV's trajectory are denoted by \( \mathbf{r}_{\text{x}}, \mathbf{r}_{\text{y}}  \in \mathbb{R}^{1 \times S} \), respectively, forming a comprehensive mapping of the UAV's spatial positions and RSS measurements. Upon completing its flight, the UAV estimates the RF source's location, \(\hat{\mathbf{P}}_{\text{RF}} = (\hat{x}, \hat{y})\), by analyzing the collected RSS measurements \(\mathbf{p}_{\text{r}}\), assuming the source is fixed at a height of \(z = 5 \, \text{m}\) and limiting the estimation to 2D positioning.

Trajectory planning can be categorized into two primary approaches: fixed waypoints and autonomous-waypoints. In the fixed-waypoint approach, the UAV follows a predefined flight path uploaded prior to deployment, systematically collecting RSS measurements along this trajectory. Conversely, the autonomous trajectory approach enables the UAV to dynamically adjust its waypoints based on observed signal strength or other experimenter-defined criteria. 

This study employs a fixed-plan trajectory due to its predictability and systematic coverage, ensuring comprehensive data collection without any regions being overlooked. The UAV operates at a constant altitude of \( r_{s, \rm z} = 30 \, \text{m} \) for all \( s \in \{1, 2, \ldots, S\} \), simplifying navigation and minimizing path deviations. The selected spiral trajectory, illustrated in Fig.~\ref{Sys} (blue solid lines), covers the overlapping area of the green and blue regions. The lower half of the blue region is excluded, as it contains no RF sources, and flying there will be of no use. This design allows efficient RSS data collection within the given operational time constraints.



\begin{figure}
	\includegraphics[width=0.8\linewidth]{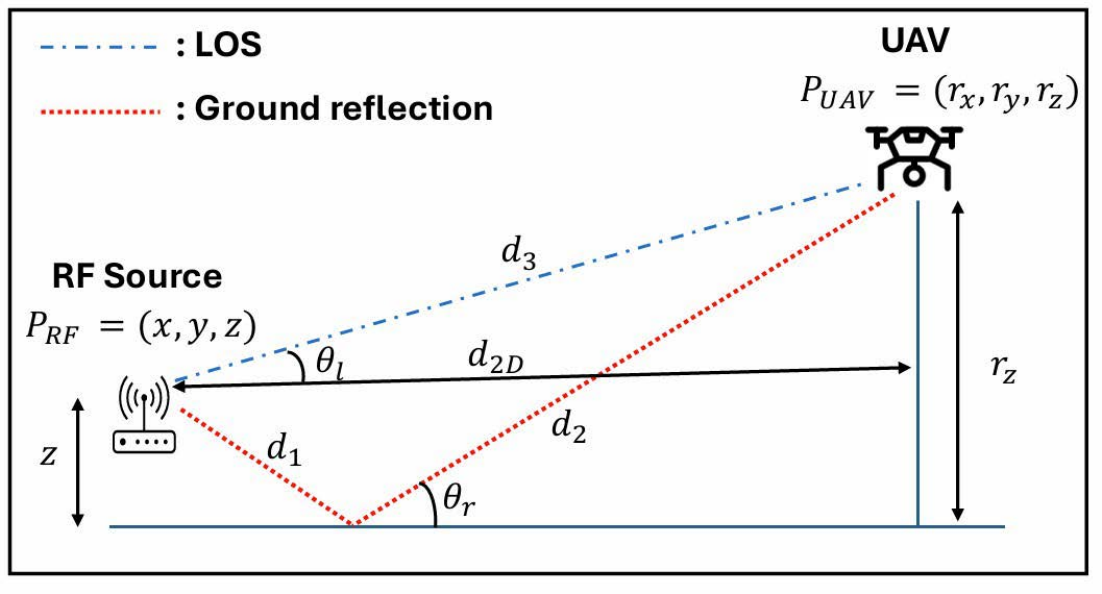}
	\centering
	\caption{Illustration of the two-ray model, showing the direct and reflected signal paths between the transmitter (RF source) and receiver (UAV).}
	\label{Tworay}
\vspace{-0.85cm}
\end{figure} 

\subsection{Two Ray  Model} \label{tworaymodel}
The Free Space Path Loss (FSPL) model, widely used for A2G propagation in line-of-sight (LoS), is simple but ignores multipath and environmental factors \cite{10605607}. To address this, we build upon the two-tay A2G model which describes the RSS as the combined effect of the LoS and reflected ray, and is mathematically expressed as \cite{maeng2023impact}:

\begin{equation}
\begin{split} \label{eq:2Ray}
p_{2 \rm ray} (\theta,\phi) = P_{\rm t} \left( \frac{\lambda}{4 \pi} \right)^2 
\Bigg| 
\frac{\sqrt{G_{\rm tx}(\phi_{\rm l},\theta_{\rm l})G_{\rm rx}(\phi_{\rm l},\theta_{\rm l})}}{d_{3}} \\
+ \Gamma(\theta_{\rm r}) 
\frac{\sqrt{G_{\rm tx}(\phi_r,\theta_r)G_{\rm rx}(\phi_{\rm r},\theta_{\rm r})} 
e^{-j \Delta \psi_{\rm phase}}}{d_{1}+d_2} 
\Bigg|^2,
\end{split}
\end{equation}

where \( p_{2 \rm ray} \) represents the RSS, \( P_t \) denotes the transmitted signal power, and \( \lambda \) is the wavelength of the transmitted signal. \( \theta_l = \mathsf{tan}^{-1}\left(\frac{z- r_{\rm z}}{d_{\rm 2D}}\right) \) and \( \theta_r = \mathsf{tan}^{-1}\left(\frac{z + r_{\rm z}}{d_{\rm 2D}}\right) \) are the elevation angles for the LoS and reflected rays, respectively. The distances \(d_1, d_2, d_3\) are defined in Fig. \ref{Tworay}. Additionally, \( \phi_{\rm l} \) and \( \phi_{\rm r} \) denote the azimuth angles for the LoS and reflected rays, respectively, while \( G_{\rm tx} \) and \( G_{\rm rx} \) represent the transmitter and receiver antenna gains as functions of \( \theta \) and \( \phi \). These gains encapsulate the directional characteristics of the antennas and their dependence on the geometric orientation of propagation paths. \( \psi_{\rm phase} \) represents the phase difference between the two propagation paths, and \(\Gamma(\theta_{\rm r})\) denotes the ground reflection coefficient~\cite{maeng2023impact}.



Antenna gain, a critical factor in A2G propagation models, directly influences the RSS. Two approaches are employed to evaluate its impact on the A2G propagation model. The first utilizes an elevation-angle-dependent \textquotedblleft donut-shaped\textquotedblright\  dipole antenna pattern, with the expressions for \( G_{\text{tx}} \) and \( G_{\text{rx}} \) provided in \cite{kwon2023rf}. The second approach utilizes measured 3D antenna patterns obtained through anechoic chamber measurements. The transmitter and receiver antenna (RM-WB1-DN) pattern is sourced from the IEEE DataPort repository \cite{maeng20233d}. As emphasized in \cite{maeng2023impact}, incorporating realistic antenna patterns significantly enhances the accuracy of the two-ray model, achieving closer alignment with measured RSS.
\vspace{-0.2cm}
\subsection{\emph{Enhanced Two-Ray} Model}
The conventional two-ray model, commonly used in A2G communication, captures RSS variations but fails to account for UAV dynamics, including roll (\(\Phi_{\rm r}\)), pitch (\(\Phi_{\rm p}\)), and yaw (\(\Phi_{\rm y}\)), as well as shadowing from the UAV's structural components (i.e., legs). These limitations hinder its ability to accurately model the complexities of the A2G propagation environment. In particular, \(\Phi_{\rm r}\), \(\Phi_{\rm p}\), and \(\Phi_{\rm y}\) represent the UAV's rotational dynamics around its longitudinal (x, nose-to-tail), lateral (y, side-to-side), and vertical (z) axes, respectively. These rotations alter the antenna's \(\theta\) (elevation) and \(\phi\) (azimuth) angles relative to the receiver, directly impacting antenna gain and LOS conditions.  Specifically, roll modifies the azimuth by tilting the main lobe sideways, pitch adjusts the elevation by tilting the lobe upward or downward, and yaw alters the azimuth by rotating the antenna horizontally. Accurate modeling of the A2G propagation model requires accounting for these orientation changes, particularly for UAVs equipped with highly directional antennas.

The objective is to update the \(\theta\) and \(\phi\) angles defining the antenna's direction relative to the receiver after UAV movements (\(\Phi_{\rm r}\), \(\Phi_{\rm p}\), \(\Phi_{\rm y}\)). This is achieved by applying the combined rotation matrix to the initial direction vector, accounting for angular shifts in 3D space. This transformation ensures the vector reflects the new orientation, enabling accurate alignment for precise antenna gain computation.

The antenna's initial pointing direction, denoted as \(\mathbf{a}\), prior to any UAV-induced movement, is expressed as:  
\begin{equation}
    \begin{aligned}
        \mathbf{a} = \Big[\mathsf{cos}(\theta)\mathsf{cos}(\phi),\; 
        \mathsf{cos}(\theta)\mathsf{sin}(\phi),\; 
        \mathsf{sin}(\theta)\Big]^T~.
    \end{aligned}
\end{equation}


\begin{figure*}[t]
\small
    \centering
    \begin{equation}
    \begin{aligned}
        \mathbf{R}_{\rm r}(\Phi_{\rm r}) &=
        \begin{bmatrix}
        1 & 0 & 0 \\
        0 & \mathsf{cos}(\Phi_{\rm r}) & -\mathsf{sin}(\Phi_{\rm r}) \\
        0 & \mathsf{sin}(\Phi_{\rm r}) & \mathsf{cos}(\Phi_{\rm r})
        \end{bmatrix}
        \;,\;
        \mathbf{R}_{\rm p}(\Phi_{\rm p}) = 
        \begin{bmatrix}
        \mathsf{cos}(\Phi_{\rm p}) & 0 & \mathsf{sin}(\Phi_{\rm p}) \\
        0 & 1 & 0 \\
        -\mathsf{sin}(\Phi_{\rm p}) & 0 & \mathsf{cos}(\Phi_{\rm p})
        \end{bmatrix}
        \;,\;
        \mathbf{R}_y(\Phi_{\rm y}) = 
        \begin{bmatrix}
        \mathsf{cos}(\Phi_{\rm y}) & -\mathsf{sin}(\Phi_{\rm y}) & 0 \\
        \mathsf{sin}(\Phi_{\rm y}) & \mathsf{cos}(\Phi_{\rm y}) & 0 \\
        0 & 0 & 1
        \end{bmatrix}~.
    \end{aligned}
    \label{eq:rotation_matrices}
    \end{equation}
    \noindent\rule{\textwidth}{0.2pt}  
    \vspace{-5mm}  
\end{figure*}

The rotation matrices corresponding to $\Phi_{\rm r}$, $\Phi_{\rm p}$, $\Phi_{\rm y}$ are denoted by $\mathbf{R}_{\rm r}$, $\mathbf{R}_{\rm p}$, and $\mathbf{R}_{\rm y}$, respectively, as defined in \eqref{eq:rotation_matrices}.
The overall rotation matrix \(\mathbf{R}\), representing the combined transformation from \(\Phi_{\rm r}\), \(\Phi_{\rm p}\), and \(\Phi_{\rm y}\), is obtained by sequentially multiplying the individual rotation matrices:
\begin{equation}
    \begin{aligned}
        \mathbf{R} =  \mathbf{R}_{\rm r}(\Phi_{\rm r}) \mathbf{R}_{\rm p}(\Phi_{\rm p}) \mathbf{R}_{\rm y}(\Phi_{\rm y})~.
    \end{aligned}
\end{equation}
The updated pointing direction of the antenna, \(\mathbf{a}'\), which represents its orientation relative to the receiver after UAV rotations is obtained by applying the overall rotation matrix \(\mathbf{R}\) to the initial antenna orientation vector \(\mathbf{a}\):
\begin{equation}
\begin{aligned}
    \mathbf{a}' = \mathbf{R} \mathbf{a} ~.
    \end{aligned}
\end{equation}  
After deriving the updated direction vector \(\mathbf{a}'\) in Cartesian coordinates, it is converted back to the updated angles \(\theta'\) and \(\phi'\) as follows:
\begin{equation}
    \begin{aligned}
        \theta' &= \mathsf{sin}^{-1}\left(\mathbf{a}'(3)\right), \quad
        \phi' &= \mathsf{tan}^{-1}\left(\mathbf{a}'(2),\mathbf{a}'(1)\right)~.
    \end{aligned}
\end{equation}

The updated angles \(\theta'\) and \(\phi'\) are utilized to compute the antenna gain with the two-ray model (Subsection \ref{tworaymodel}). This transformation ensures that the effects of UAV orientation changes are accurately incorporated, allowing for precise modeling of the effective antenna gains.


To further enhance the accuracy of the two-ray model, shadowing effects from the UAV's legs are incorporated. The UAV body is represented as a hexagon with \(L = 3\) legs extending downward at angles \( \psi_{\ell} \) (\(\ell \in \{1, 2, 3\}\)), where \( \psi_1 = 39^\circ \), \( \psi_2 = 150^\circ \), and \( \psi_3 = 270^\circ \). These legs obstruct signals arriving at their respective angles, significantly reducing the RSS.

The enhanced two-ray model incorporates UAV angles (\(\Phi_{\rm r}\), \(\Phi_{\rm p}\), \(\Phi_{\rm y}\)) and body shadowing effects. The RSS, denoted as \(p_{\rm E-2ray}\), is given by:
\begin{equation} \label{Etworay}
p_{\rm E-2ray}(\theta',\phi') \doteq \begin{cases}
p_{2 \rm ray} (\theta',\phi') - \Upsilon, & \text{if }  S=1 \\
p_{2 \rm ray} (\theta',\phi'), & \text{otherwise}
\end{cases}~,
\end{equation}
where, \( p_{2 \rm ray} (\theta', \phi') \) denotes the RSS from the conventional two-ray model, calculated using the updated elevation (\(\theta'\)) and azimuth (\(\phi'\)) angles. The term \(\Upsilon\) represents the shadowing loss subtracted from the RSS when body-induced shadowing is detected (\(S = 1\)) and is defined as: 
\begin{equation}
    \begin{aligned}
       \Upsilon = 10 \beta \log_{10} \left( \frac{d_{\text{sh}}}{d_0} \right)~,
    \end{aligned}
\end{equation}

where \(d_{\text{sh}}\) is the \(d_3\) distance at which shadowing occurs, $\beta $ is the shadowing exponent, and \(d_0\) is the reference distance.

Shadowing of the transmitter is likely to occur if its azimuth angle \( \phi \) falls within any of the leg-induced shadowing intervals, defined as \( \phi \in [\psi_\ell - \Delta, \psi_\ell + \Delta] \), where \( \Delta \) represents the angular spread (in degrees) that accounts for the finite width of the leg and its surrounding obstruction.
An elevation angle check is incorporated to enhance shadowing detection, with shadowing considered only if \(\theta < \delta\), where \(\delta\) is a threshold angle (in degrees). This parameter acts as a critical cutoff to differentiate scenarios where shadowing is caused by obstructions from those where the transmitter is either directly beneath the UAV or at a high enough elevation angle to maintain LOS. In summary, the transmitter is considered shadowed if:
\begin{equation}
S \doteq \begin{cases}
1, & \text{if }  \phi \in [\psi_l - \Delta, \psi_l + \Delta] \text{ and } \theta < \delta, \\
0, & \text{otherwise.}
\end{cases}
\end{equation}

This combined azimuth-elevation approach effectively captures realistic scenarios where the UAV structure obstructs signals, particularly when the transmitter is positioned near the horizon.

\begin{figure}
	\includegraphics[width=0.99\linewidth]{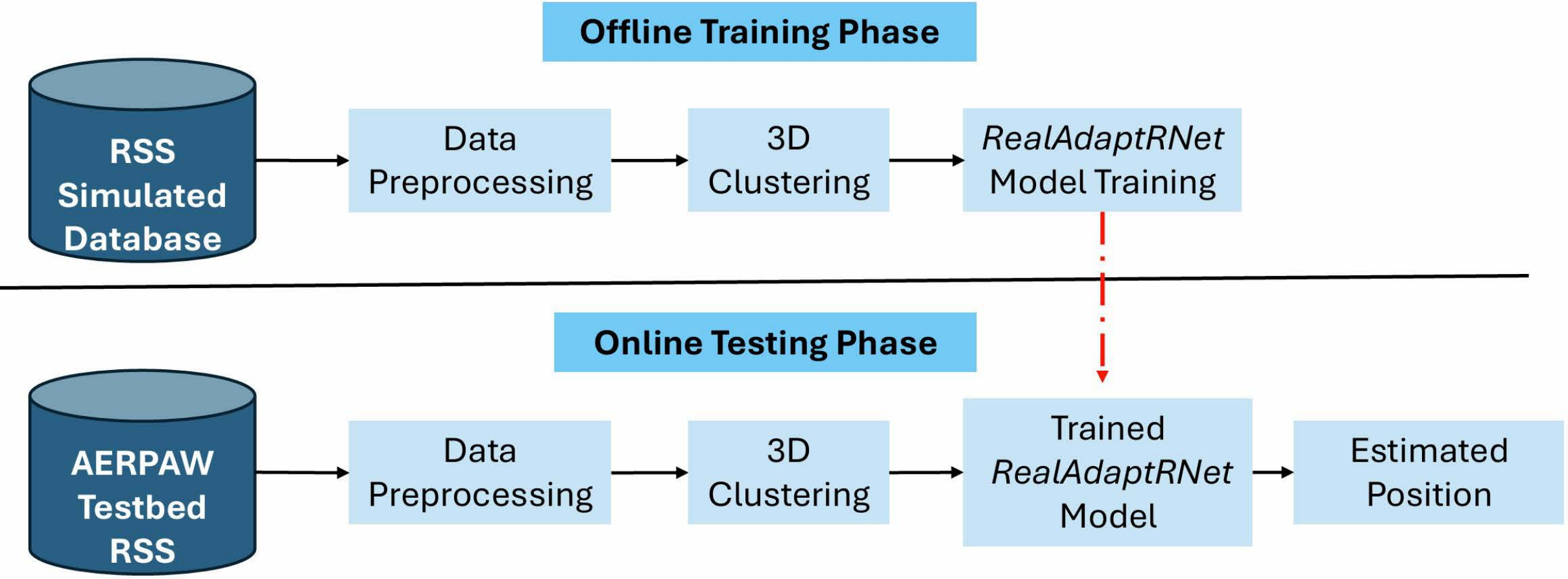}
	\centering
	\caption{DL-based pipeline for RFSL.}
	\label{DLPipeline}
\vspace{-0.4cm}
\end{figure} 


\begin{figure*}[t] 
    \centering
    \begin{subfigure}[b]{0.4\textwidth}
        \centering
        \includegraphics[width=\textwidth]{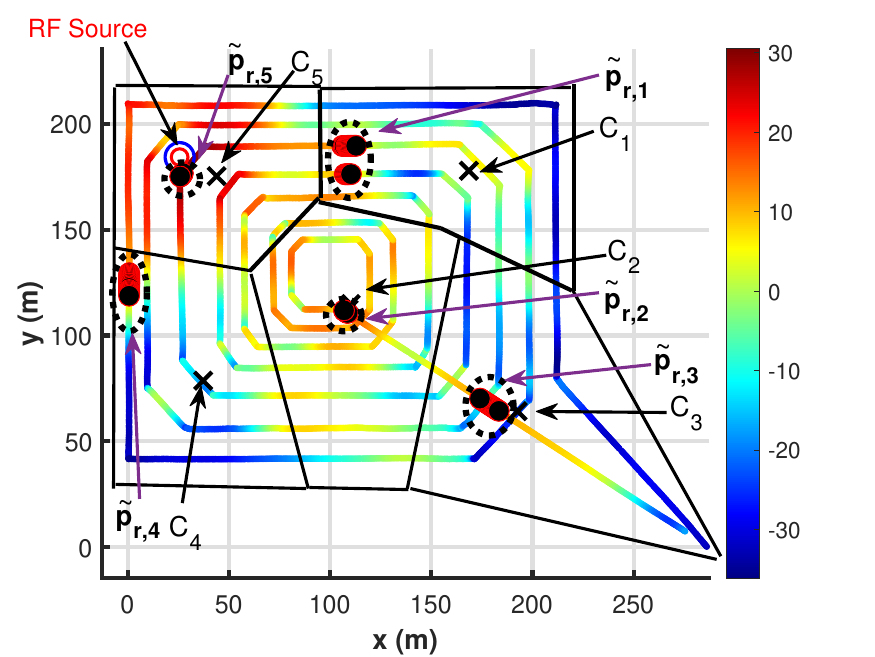}
        \caption{Data clustering of RSS values along UAV trajectory.}
        \label{Clustering}
    \end{subfigure}
    \quad
    \begin{subfigure}[b]{0.55\textwidth}
        \centering
        \includegraphics[width=\textwidth]{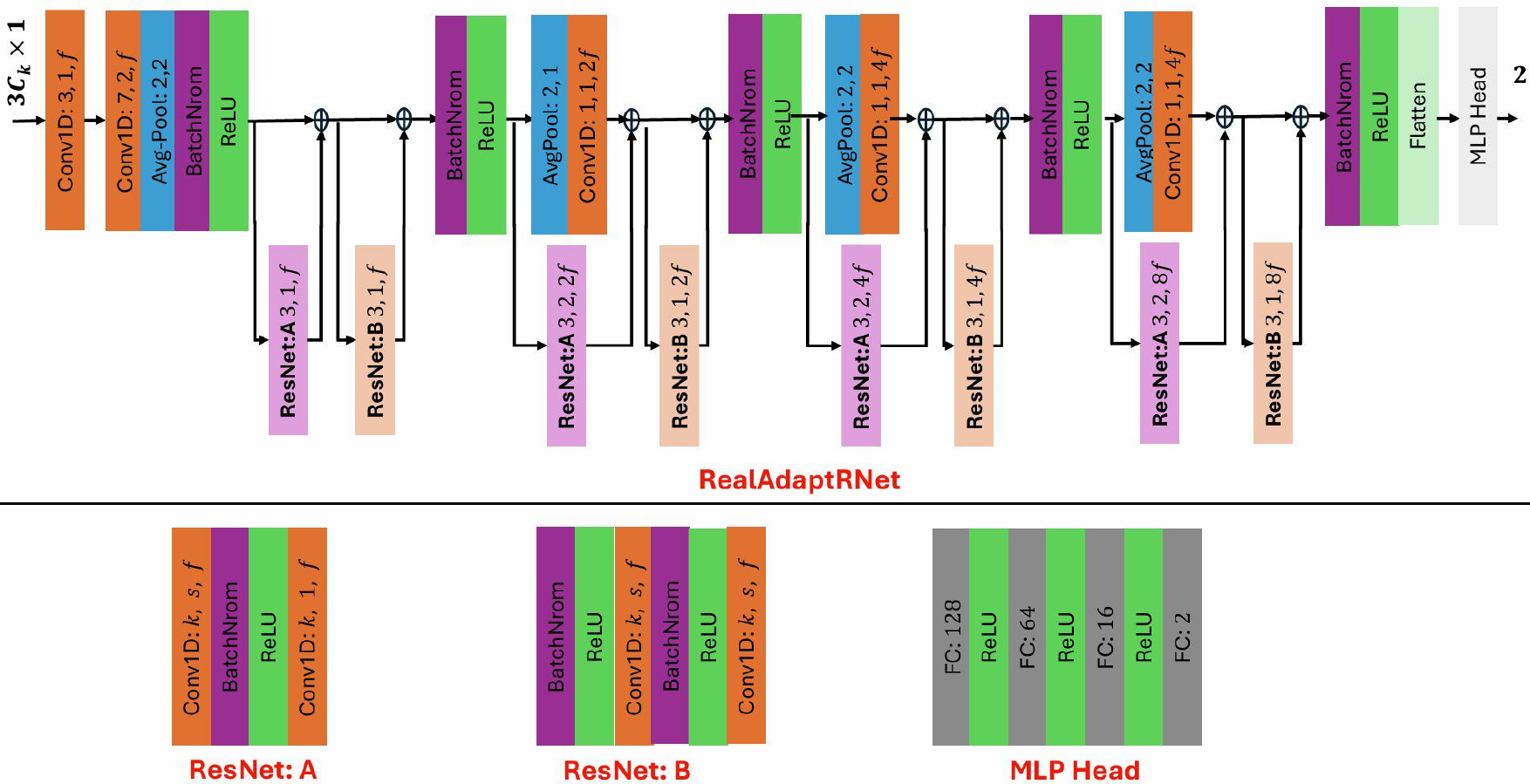}
        \caption{Proposed \emph{RealAdaptRNet} architecture.}
        \label{Resnet}
    \end{subfigure}
   
    \caption{(a) The received RSS values collected using the AERPAW testbed along the UAV trajectory for an RF source positioned at $(x, y) = (20, 184)$. Five clusters are considered, with black boundaries indicating the regions where points belong to their respective clusters. (b) The proposed \emph{RealAdaptRNet} uses 1D convolutional layers denoted as Conv1D: \( k, s, f \), where \( k \) is the kernel size, \( s \) is the stride, and \( f \) is the number of filters. In ResNet Block A, the last convolutional layer always uses a stride of 1, while the other parameters depend on the placement within the architecture.}
    \label{fig:ViT_Model}
    \vspace{-5mm} 
\end{figure*}

\section{DL-Based Model For RF Localization with UAV}
This section presents a DL-based approach for RFSL, utilizing RSS measurements collected by the UAV during its flight. The proposed approach is trained on simulated datasets generated using the A2G models described in Section \ref{sec:system} and validated in real-world scenarios using the AERPAW testbed. Although the \emph{enhanced two-ray} model improves the accuracy of simulations, it cannot fully replicate the complexities of real-world environments. Consequently, a simple DL model may struggle to generalize effectively. To address this, we propose the \emph{3D Clustering-Based RealAdaptRNet} model, designed to overcome the sim2real gap and enhance localization performance. 

The overall DL pipeline for RFSL is depicted in Fig. \ref{DLPipeline}. The pipeline operates in two phases: (1) Offline Training Phase and (2) Online Testing Phase. In the offline training phase, the proposed \emph{RealAdaptRNet} model is trained on datasets generated from the A2G propagation models described in Section \ref{sec:system}. For the A2G propagation model, a dataset of \(N\) data points was generated by randomly placing the RF source at \(N\) distinct locations. For each RF location, the UAV completed its spiral trajectory, collecting RSS measurements to form the dataset. Each data-point, denoted as $\mathbf{p}_{\text{r}, n}= [p^1_{\rm E-2ray}, \cdots, p^{S}_{\rm E-2ray}]^T \in \mathbf{R}^{(S\times 1)}$, represents the RSS measurements for the \(n\)-th RF source location generated using the enhanced two-ray (E-2ray) model\footnote{This data point can also be generated using the free-space or two-ray models, depending on the propagation scenario being simulated.}, where $n \in (1, \cdots, N)$, while the output is the estimated 2D coordinates of the RF source, $P_{\rm RF, n} \in \mathbf{R}^{(1 \times 2)} $. The online phase involves deploying the trained \emph{RealAdaptRNet} model in the AERPAW testbed for real-world localization. 

\vspace{-0.3cm}
\subsection{Data Preprocessing}
UAV during its trajectory collects roughly \(S = 20k\) RSS samples, directly using such a large number of samples for DL-based localization significantly increases the model’s computational complexity in terms of FLOPs and parameters. To address this, and considering that RSS values often exhibit deep fades, we group the RSS samples $\mathbf{p}_{\text{r}, n}$ into non-overlapping segments of size \(S_{\rm g}\). To reduce the data size and smooth the RSS signals while preserving key information, the average RSS value is computed for each group, with the resulting output denoted as \( \mathbf{\check{p}}_{\text{r}, n} \in \mathbf{R}^{1\times S/N_{\text{g}}}\). 

RSS measurements are smoothed using discrete convolution with a Gaussian kernel \(\mathbf{g}\):
\begin{equation}
    \begin{aligned}
        \mathbf{g}(\mu) \;=\; \frac{1}{\sqrt{2\pi \,\sigma^2}} \,\exp\!\Bigl(-\frac{\mu^2}{2\,\sigma^2}\Bigr) ~,
    \end{aligned}
\end{equation}
where \(\mu\) ranges from \(-\lfloor k_s/2 \rfloor\) to \(\lfloor k_s/2 \rfloor\) and \(k_s = 6\,\sigma + 1\). The kernel is normalized so that \(\sum_{\mu} \mathbf{g}(\mu) = 1\). The smoothed RSS values \(\hat{\mathbf{p}}_{\rm r, n}\) are then obtained by convolving the raw measurements \(\check{\mathbf{p}}_{\rm r, n}\) with \(\mathbf{g}(\mu)\) as:  
\begin{equation}
    \begin{aligned}
        \hat{\mathbf{p}}_{\rm r, n}[i] \;=\; \sum_{\mu=-\lfloor k_s/2 \rfloor}^{\lfloor k_s/2 \rfloor} \check{\mathbf{p}}_{r}^n[i-\mu] \, \mathbf{g}(\mu).
    \end{aligned}
\end{equation}
 
This filtering step reduces high-frequency noise while preserving the underlying signal trend.

As a baseline, we also considered using the \emph{Normalized RSS}, denoted as \(\check{\mathbf{p}}_{\text{r}, n}\), as input to the \emph{RealAdaptRNet} model (to be defined in Subsection 
\ref{DLModel}) for localization. Gaussian filtering is excluded in this setup since the model's convolutional layers inherently perform the smoothing operation.
\vspace{-0.3cm}
\subsection{3D Clustering}

After Gaussian smoothing, the processed data for the \(n\)-th RF source location is represented as \(\mathcal{D}_{n} = [\mathbf{r}_{\text{x}, n}, \mathbf{r}_{\text{y}, n}, \hat{\mathbf{p}}_{\text{r}, n}]\). A 3D clustering approach is then applied using k-nearest neighbors (KNN) on \(D_n\), partitioning the data into clusters \(C_k\), where black lines differentiate samples belonging to each cluster. While the clustering is depicted for visualization purposes in 2D, the actual clustering is performed in 3D space, incorporating both spatial coordinates and RSS values (\(\hat{\mathbf{p}}_{\text{r}, n}\)) to effectively capture the spatial and signal characteristics of the UAV's trajectory.

Unlike traditional clustering based solely on RSS values, which may group distant points with similar signal levels into the same cluster, the 3D clustering strategy effectively captures both the spatial and signal characteristics of the UAV's trajectory. This prevents the merging of spatially distant but signal-similar points, which can occur due to multipath effects or shadowing. 

For each cluster \( C_c \), where \( c \in \{1, \ldots, T_{\rm c}\} \), the \( n_{\text{t}} \) samples with the highest RSS values are identified and denoted as \( \mathbf{\tilde{p}}_{r, n, c} = \max(\hat{\mathbf{p}}_{\text{r}, n, c}, n_{\text{t}}) \), where \( \hat{\mathbf{p}}_{\text{r}, n, c} \) represents the complete set of RSS measurements associated with cluster \( C_c \). The corresponding coordinates \( (\tilde{\mathbf{r}}_{\text{x}, n, c}, \tilde{\mathbf{r}}_{\text{y}, n, c}) \) associated with the highest RSS values are similarly selected. These selected samples are depicted in circles within each cluster in Fig. \ref{Clustering}.  This process forms regions reflecting different proximities to the RF source—some near the source with high RSS values and others farther away with lower RSS values. This ensures the selected high-RSS samples provide a well-distributed representation of the UAV's trajectory, avoiding bias toward densely sampled areas or specific points. The average RSS for the selected highest values is computed as \( \bar{p}_{\text{r}, n, c} = \frac{1}{n_{\text{t}}} \sum_{j=1}^{n_{\text{t}}} \mathbf{\tilde{p}}_{\text{r}, n, c}(j) \). Similarly, the corresponding mean coordinates are denoted by \( (\bar{r}_{\text{x}, n, c}, \bar{r}_{\text{y}, n, c}) \).
By prioritizing high-RSS samples within spatially coherent clusters, and taking the average, the proposed approach mitigates the effects of multipath fading and reduces sensitivity to outliers.

The vector $\Ddot{\mathbf{p}}_{\rm r, n}$, called the \emph{3D Clustering} input, is formed by concatenating average RSS values with centroid coordinates of all clusters and serves as input to the \emph{RealAdaptRNet} model. Formally, it is represented as:
\begin{equation}
    \begin{aligned}
        \Ddot{\mathbf{p}}_{\rm r, n} = \left[\bar{p}_{\text{r},1}^{n} , \bar{r}_{\text{x},1}^{n}, \bar{r}_{\text{y},1}^{n},  \cdots, \bar{p}_{\text{r},C_{\rm k}}^{n} , \bar{r}_{\text{x},C_{\rm k}}^{n}, \bar{r}_{\text{y},C_{\rm k}}^{n}\right]^T \in \mathbb{R}^{3C_{k} \times 1}~.
    \end{aligned}
\end{equation}
This input vector provides a structured and concise representation of the key spatial and signal characteristics for the RFSL task. Using 3D clustering offers several advantages over directly inputting all RSS values $\mathbf{\check{p}}_{\text{r}, n}$ into the model for localization. The proposed clustering approach reduces the input size significantly, thereby lowering computational complexity and memory usage. Additionally, it improves the model's robustness by summarizing key signal patterns.

\vspace{-0.2cm}
\subsection{\emph{RealAdaptRNet} Model} \label{DLModel}

Our proposed \emph{RealAdaptRNet} model as shown in Fig. \ref{Resnet} leverages a hybrid 1D ResNet-MLP architecture designed to estimate precise RF source position. The one-dimensional input (i.e., \( \Ddot{\mathbf{p}}_{\text{r}, n} \in \mathbb{R}^{3C_k \times 1}, \text{or},   
 \check{\mathbf{p}}_{\text{r}, n} \in \mathbf{R}^{S/N_{\text{g}} \times 1} \)) is transformed by the first convolutional layer into an expanded feature space of dimension \( \mathbb{R}^{N_{\text{inp}} \times f} \), where \( f \) represents the number of filters in the convolutional layer, and \(N_{\text{inp}}\) in the input size of one data samples (i.e., \(N_{\text{inp}}=3C_k\) for 3D clustering approach). This transformation projects the input into a multi-channel representation, allowing the network to capture diverse and complex feature patterns crucial for spatial feature extraction. The network is composed of multiple convolutional blocks with residual connections to preserve information and stabilize gradient flow. Each block consists of 1D convolutions (kernel sizes of $k=3$ or $k=1$), batch normalization, ReLU activations, and average pooling for effective feature extraction and dimensionality reduction. The MLP head consists of \( N_{\text{MLP}}=4 \) fully connected layers that progressively reduce the feature space, outputting the predicted 2D coordinates.

\begin{figure}
	\includegraphics[width=0.8\linewidth]{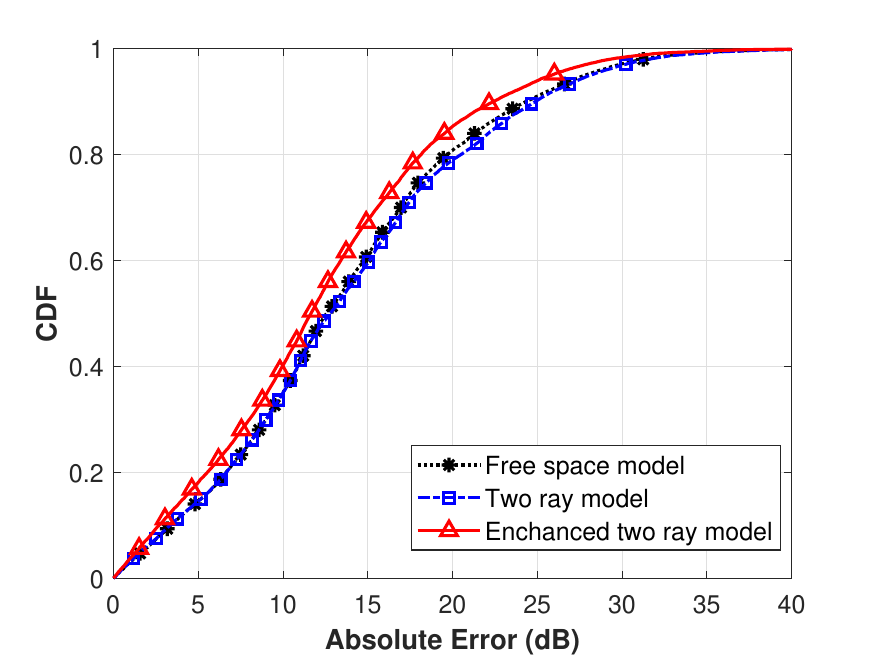}
	\centering
	\caption{CDF of absolute error (in dB) for the free-space, two-ray, and enhanced two-ray models compared to real-world RSS data. }
	\label{Abserror}
\vspace{-0.4cm}
\end{figure}

\begin{figure*}[t] 
    \centering
    \begin{subfigure}[b]{0.31\textwidth}
        \centering
        \includegraphics[width=\textwidth]{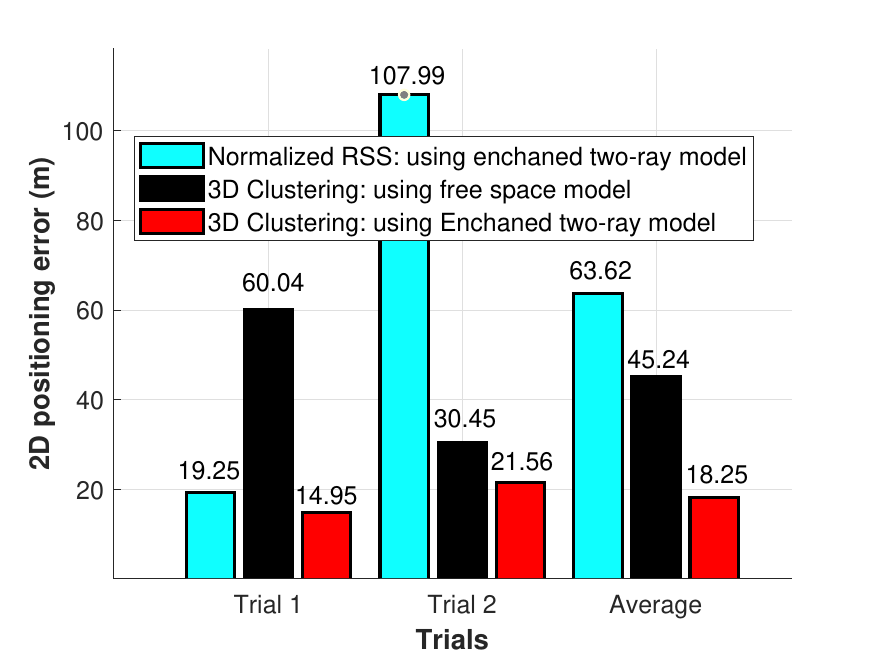}
        \caption{2D positioning error comparison across trials for \emph{RealAdaptRNet} using normalized RSS and clustering-based models.}
        \label{fig:CA1}
    \end{subfigure}
    \quad
    \begin{subfigure}[b]{0.31\textwidth}
        \centering
        \includegraphics[width=\textwidth]{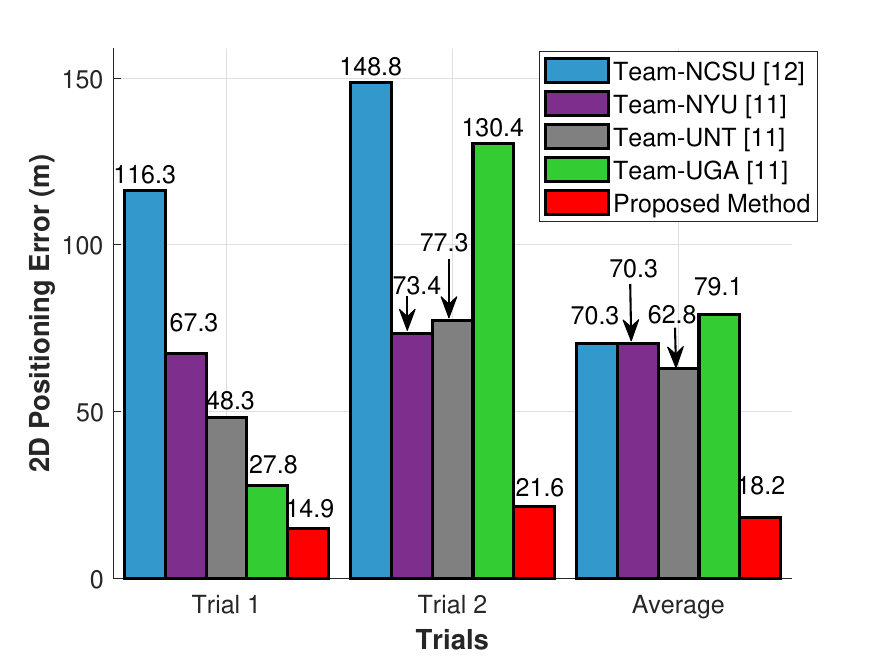}
        \caption{Comparison of proposed method (3D Clustering: using Enchaned two-ray model) with other teams in the AFAR.}
        \label{fig:CA2}
    \end{subfigure}
    \quad
    \begin{subfigure}[b]{0.31\textwidth}
        \centering
        \includegraphics[width=\textwidth]{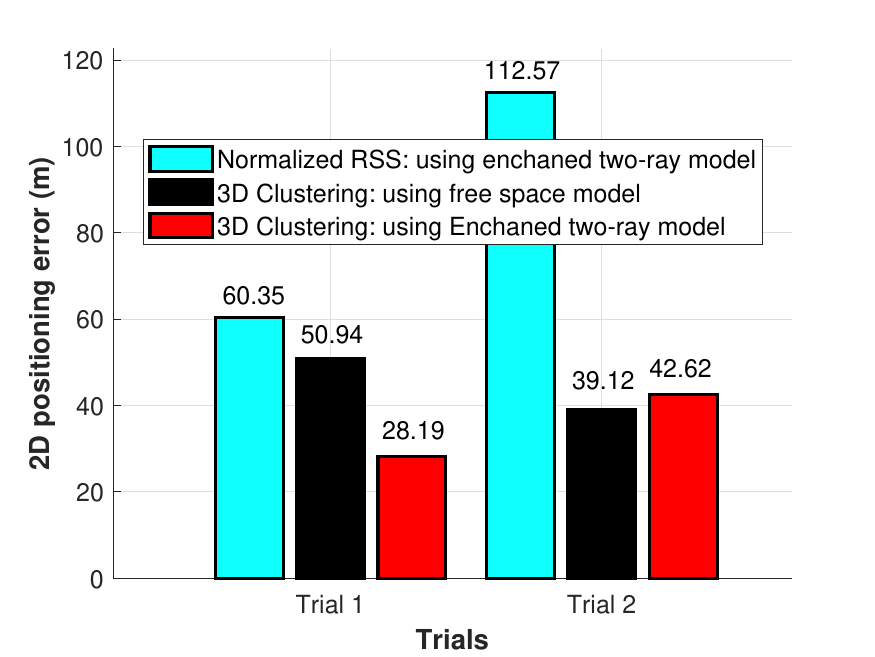}
        \caption{Generalization performance across UAV trajectories using the proposed method and baselines.}
        \label{fig:CA3}
    \end{subfigure}
    \caption{Analysis of 2D positioning errors across trials, comparisons with AFAR challenge results, and evaluations of UAV trajectory performance for the proposed \emph{3D Cluster-Based RealAdaptRNet} model and baseline approaches.}
    \label{fig:CA}
    \vspace{-5mm} 
\end{figure*}


\section{Numerical Results and Analysis}\label{sec:Nume}

In the numerical analysis, the transmit power and central frequency were set to \( P_{\text{t}} = 41 \, \text{dBm} \) and \( f_{\text{c}} = 3.32 \, \text{GHz} \), respectively, with a relative permittivity of \( \epsilon_0 = 2 \). To simulate UAV movement effects, the roll (\( \Phi_{\text{r}} \)), pitch (\( \Phi_{\text{p}} \)), and yaw (\( \Phi_{\text{y}} \)) angles were uniformly sampled from \( \mathcal{U}(-5^\circ, 5^\circ) \). The angular spread was set to \( \Delta = 5^\circ \), the shadowing threshold elevation angle to \( \delta = 10^\circ \), and the reference distance to \( d_0 = 1 \, \text{m} \). Gaussian noise with zero mean and unit variance is added to the RSS for training, and \(\Upsilon\) is obtained using the linear least squares method. To reduce data dimensionality, a grouping factor of \( N_{\text{g}} = 2 \) was applied, halving the number of samples. Gaussian smoothing with a kernel (\( \sigma = 20 \)) mitigated high-frequency noise while preserving signal characteristics. The RSS values were clustered into \( C_k = 20 \) clusters, and the top \( n_{\text{t}} = 40 \) RSS samples from each cluster were selected to construct the feature vectors. The proposed \emph{RealAdaptRNet} algorithm is implemented using Jax and Flax libraries.
\vspace{-0.2cm}
\subsection{Analysis of Enchaned Two Ray Path Loss Model}

This section evaluates the performance of the proposed \emph{enhanced two-ray model} against the two-ray and free-space propagation models. Fig.~\ref{Abserror} illustrates the absolute error (dB) between the measured real-world RSS and the predictions from the free-space,  two-ray with \textquotedblleft donut-shaped\textquotedblright\ gains, and \emph{enhanced two-ray} propagation models. The free-space and two-ray models exhibit comparable performance; however, the two-ray model occasionally performs worse due to the oversimplified antenna gain assumptions. In contrast, the enhanced two-ray model demonstrates superior accuracy, with over 90\% of the errors contained within 22 dB, outperforming both the free-space and standard two-ray models. By integrating antenna directionality through the angles \( \Phi_{\text{r}}, \Phi_{\text{p}}, \Phi_{\text{y}} \) and accounting for UAV body-induced shadowing effects, the model effectively captures real-world RSS variations, making it highly reliable for realistic propagation scenarios.

\vspace{-0.2cm}
\subsection{Analysis of \emph{RealAdaptRNet} model for RFSL}
This subsection evaluates the real-world localization performance of the proposed \emph{3D Cluster-Based RealAdaptRNet} model, focusing on its ability to generalize from simulated datasets to real-world scenarios. Results on the simulated dataset are omitted due to space constraints. Experimental trials were conducted using the AERPAW testbed, with the RF source placed at two distinct locations for validation, referred to as trials. In Trial 1, the RF source was positioned at \(P_{\rm RF} = (17.67, 194.29, 5)\,\mathrm{m}\). For Trial 2, to evaluate the robustness of the proposed method, the RF source was placed outside the overlap region, within the green area depicted in Fig.~\ref{Sys}, at \(P_{\rm RF} = (206, 275.80, 5)\,\mathrm{m}\), as illustrated in Fig.~\ref{Sys}.

\subsubsection{\textbf{Performance Evaluation of 3D Clustering and Propagation Models}}
In this subsection, the performance of the \emph{RealAdaptRNet} model is evaluated using the enhanced two-ray propagation model with two input representations: the proposed 3D clustering approach (\(\Ddot{\mathbf{p}}_{\rm r}^n\)) and the normalized RSS approach (\(\mathbf{\check{p}}_{\rm r}^n\)). Furthermore, the model trained with the enhanced two-ray propagation model is compared to its counterpart trained with the free-space model, both utilizing the 3D clustering approach, to assess the impact of different propagation models on localization performance.

Fig.~\ref{fig:CA1} presents the 2D positioning errors across trials, along with the average errors. The \emph{RealAdaptRNet} model trained on the proposed \emph{3D clustering} approach using the \emph{enhanced two-ray} propagation model consistently outperforms all other approaches. In Trial 1, the \emph{RealAdaptRNet} model using normalized RSS input trained on the \emph{enhanced two-ray} model performed well as it outperformed the model using 3D clustering input trained on the free-space model. This result highlights that the \emph{free-space} model fails to accurately represent real-life scenarios, as it primarily emphasizes high RSS values near the RF source, whereas real-world conditions exhibit high RSS values over a much broader region around the source, leading to confusing the \emph{RealAdaptRNet} model. For Trial 2, the error is high as the RF source is far from the closest UAV waypoint, causing the normalized RSS method to fail, while the clustering approach performs well. The \emph{RealAdaptRNet} model with normalized RSS input struggles with high RSS values at waypoints relatively near the RF, causing confusion and inaccurate predictions for Trial 2. The 3D clustering approach with the \emph{enhanced two-ray} model significantly reduces error, dropping positioning error in Trial 2 from 107.9 m (normalized RSS) to 21.56 m (3D clustering).

Moreover, the normalized RSS approach increases the \emph{RealAdaptRNet} model computational complexity, requiring 331.98M FLOPs and 6.7M parameters, compared to the 3D clustering approach's 1.3M FLOPs and 0.2M parameters—representing~255.37× fewer FLOPs and~33.5× fewer parameters. Despite this, the clustering approach with \emph{enhanced two-ray} A2G model delivers superior performance across all trials. 

Figure~\ref{fig:CA2} compares the 2D positioning error of the proposed \emph{3D Cluster-Based RealAdaptRNet} using the \emph{Enhanced Two-Ray} model with the top four finalist teams from the AFAR challenge~\cite{kudyba2024uav}, \cite{aerpaw_afarchallenge}. The proposed approach significantly outperformed all teams across trials, including Trial 2, where most teams failed to achieve low errors, demonstrating its robustness and superior performance.
\subsubsection{\textbf{Generalization Performance for Different Trajectories}}
To evaluate the generalization ability of the proposed approach, we tested the model trained on the proposed spiral trajectory using four additional trajectories from the top 4 teams in the AFAR challenge, each distinct from the training trajectory. The evaluation used the same 2 trial locations for all five trajectories (including the proposed trajectory). Fig.~\ref{fig:CA3} presents the average positioning error across all trials for the five trajectories. The normalized RSS approach exhibits consistently higher errors across all trials, as it fails to account for the changes in UAV sampling positions introduced by varying trajectories. In contrast, the proposed clustering approach, trained on both the free-space and \emph{enhanced two-ray} propagation models, performs significantly better by leveraging trajectory-agnostic features extracted by the proposed 3D clustering relevant to RFSL. The proposed \emph{3D Cluster-Based RealAdaptRNet} using the \emph{Enhanced Two-Ray} model generalizes so effectively that it outperformed the respective teams' algorithms, even when applied to their specific trajectories, despite being trained on different trajectory.

For Trial 2, where the RF source lies outside the UAV's trajectory, the free-space and enhanced two-ray models using the 3D clustering approach show similar performance due to limited RSS observations for this case, leading to comparable positioning errors. However, the enhanced two-ray model demonstrates superior accuracy in other trials, highlighting its effectiveness in capturing RSS variations in complex environments.


\section{Conclusion}
This study addresses the gap between simulated and real-world scenarios by introducing an enhanced two-ray propagation model that accounts for UAV dynamics' impact on antenna gains and shadowing caused by the UAV structure. While simulations cannot fully replicate real-world complexities, the proposed \emph{3D Cluster-Based RealAdaptRNet} model bridges this gap by learning from simulated datasets and delivering accurate RFSL in real-world conditions. The 3D clustering approach significantly improves localization accuracy compared to normalized RSS input, reducing the average error from 48.42 m (normalized RSS) to 47.61 m with the free-space model and further to 16.12 m with the enhanced two-ray model. The proposed method also achieves a 33.5× reduction in parameters, significantly lowering computational complexity without compromising performance. Validation on the AERPAW testbed confirms the model's robustness and generalization capabilities, achieving an average error of 35.403 m across all trials and trajectories, even when tested on trajectories different from those used during training. Future work will focus on further narrowing the sim2real gap, exploring adaptive trajectory planning, and developing advanced methods to enhance localization accuracy and efficiency.

\vspace{-0.2cm}

\bibliographystyle{IEEEtran}

\end{document}